\RequirePackage{rotating}
\documentclass[useAMS,usenatbib]{mnras}

\usepackage{ulem}
\usepackage{color}
\usepackage{graphics,graphicx}
\usepackage{times} 
\usepackage{amssymb}
\usepackage{amsmath}
\usepackage{natbib}
\usepackage{lscape}
\usepackage{url}
\usepackage{multirow}
\usepackage{ctable}
\usepackage{threeparttable}
\usepackage[labelfont=bf,labelsep=period]{caption}
\newif\ifAMStwofonts
\AMStwofontstrue

\def\xmm{{\it XMM-Newton}}

\def\hst{{\it HST}}

\def\chandra{{\it Chandra}}
\def\swift{{\it Swift}}
\def\swiftng{{\it Neil Gehrels Swift Observatory}}

\def\epicpn{{EPIC-pn}}
\def\epicmos1{{EPIC-MOS1}}
\def\epicmos2{{EPIC-MOS2}}
\def\epicmos{{EPIC-MOS}}

\def\nustar{{\it NuSTAR}}
\def\hexp{{\it HEX-P}}

\def\erosita{{\it eROSITA}}
\def\wise{{\it WISE}}

\def\deg{$^{\circ}$}

\def\pcmsq{\hbox{$\rm\thinspace cm^{-2}$}}
\def\H0{{km~s$^{-1}$~Mpc$^{-1}$}}

\def\ergpcmsqps{\hbox{$\rm\thinspace erg~cm^{-2}~s^{-1}$}}

\def\ergps{\hbox{erg~s$^{-1}$}}
\def\ergcmps{\hbox{\rm erg~cm~s$^{-1}$}}

\def\eddrat{$\lambda_{\rm{E}}$}

\def\msun{\hbox{$M_{\odot}$}}

\def\kte{$kT_{\rm{e}}$}
\def\ecut{$E_{\rm{cut}}$}

\def\lambdaCDM{$H_0$ = 70\,\H0, $\Omega_{\rm{M}} = 0.3$ and $\Omega_{\Lambda} = 0.7$}

\def\caldb{\textsc{caldb}}

\def\addascaspec{\textsc{addascaspec}}

\def\flx2xsp{\textsc{flx2xsp}}

\def\nustardas{\textsc{nustardas}}
\def\nupipeline{\textsc{nupipeline}}
\def\nuproducts{\textsc{nuproducts}}

\def\sas{\textsc{sas}}
\def\xmmselect{\textsc{xmmselect}}

\def\epchain{\textsc{epchain}}
\def\emchain{\textsc{emchain}}

\def\rmfgen{\textsc{rmfgen}}
\def\arfgen{\textsc{arfgen}}

\def\ciao{\textsc{ciao}}
\def\chandrarepro{\textsc{chandra\_repro}}
\def\edser{\textsc{edser}}
\def\specextract{\textsc{specextract}}

\def\reprojectobs{\textsc{reproject\_obs}}
\def\wavdetect{\textsc{wavdetect}}
\def\wcsmatch{\textsc{wcsmatch}}
\def\wcsupdate{\textsc{wcsupdate}}

\def\chisq{{$\chi^{2}$}}

\def\xspec{\hbox{\small XSPEC}}

\def\xstar{\textsc{xstar}}

\def\cabs{\textsc{cabs}}
\def\tbabs{\textsc{tbabs}}

\def\partcov{\textsc{partcov}}

\def\relxill{\textsc{relxill}}

\def\relxilllpcp{\textsc{relxilllp\_cp}}

\def\torus{\textsc{torus}}
\def\borus{\textsc{borus}}

\def\nthcomp{\textsc{nthcomp}}

\def\zashift{\textsc{zashift}}
\def\xclumpy{\textsc{xclumpy}}

\def\mgii{\hbox{\rm Mg\,{\small II}}}

\def\eg{{\it e.g.}}

\def\ie{{\it i.e.~\/}}

\def\la{\mathrel{\hbox{\rlap{\hbox{\lower4pt\hbox{$\sim$}}}{\raise2pt\hbox{$<$}}}}}
\def\ga{\mathrel{\hbox{\rlap{\hbox{\lower4pt\hbox{$\sim$}}}{\raise2pt\hbox{$>$}}}}}

\def\d25{D$_{25}$}
\def\nh{{$N_{\rm H}$}}

\def\Ha{{H$\alpha$}}
\def\Hb{{H$\beta$}}

\def\.25{0.25 keV\thinspace}

\def\kbol210{\rm $\kappa_{2-10}$}

\def\rg{$R_{\rm{G}}$}

\def\Rfrac{$R_{\rm{frac}}$}

\def\src{2MASS\,J1042+1641}
\def\predFcov{0.35}
\def\intLx{$2-5 \times 10^{44}$}
\def\intLsix{$4-9 \times 10^{45}$}
\def\intLftn{$5-11 \times 10^{45}$}
\def\intLbol{$4-10 \times 10^{46}$}

\title[X-ray Absorption and Reprocessing in \src]{X-ray Absorption and Reprocessing in the \textit{z} $\sim$ 2.5 Lensed Quasar \src}

\author[D.\,J. Walton et al.]
{\parbox{7.in}{D.\,J. Walton$^{1,2}$ \thanks{E-mail: d.walton4@herts.ac.uk},
M. T. Reynolds$^{3}$,
D. Stern$^{4}$,	
M. Brightman$^{5}$,
C. Lemon$^{6}$
\\[0.25cm]
\footnotesize
$^{1}$ \it{Centre for Astrophysics Research, University of Hertfordshire, College Lane, Hatfield AL10 9AB, UK} \\
$^{2}$ \it{Institute of Astronomy, University of Cambridge, Madingley Road, Cambridge CB3 0HA, UK} \\
$^{3}$ \it{Department of Astronomy, University of Michigan, 1085 S. University, Ann Arbor, MI 48109, USA} \\
$^{4}$ \it{Jet Propulsion Laboratory, California Institute of Technology, Pasadena, CA 91109, USA} \\
$^{5}$ \it{Space Radiation Laboratory, California Institute of Technology, Pasadena, CA 91125, USA} \\
$^{6}$ \it{Institute of Physics, Laboratory of Astrophysics, Ecole Polytechnique F\'ed\'erale de Lausanne (EPFL), Observatoire de Sauverny,
1290 Versoix, Switzerland}
}}
\date{}

\begin{document}
\pagerange{\pageref{firstpage}--\pageref{lastpage}}
\maketitle
\label{firstpage}

\begin{abstract}
We present new broadband X-ray observations of the $z \sim 2.5$ lensed quasar
\src, combining \xmm, \chandra\ and \nustar\ to provide coverage of the X-ray
spectrum over the 0.3--40\,keV bandpass in the observed frame, corresponding to the
$\sim$1--140\,keV band in the rest-frame of \src. The X-ray data show clear evidence
for strong (but still Compton-thin) X-ray absorption, $N_{\rm{H}} \sim 3-4 \times 
10^{23}$\,\pcmsq, in addition to significant reprocessing by Compton-thick material
that must lie away from our line-of-sight to the central X-ray source. We test two
different interpretations for the latter: first that the reprocessing occurs in a classic
AGN torus, as invoked in unification models, and second that the reprocessing occurs
in the accretion disc. Both models can successfully reproduce the observed spectra,
and both imply that the source is viewed at moderately low inclinations ($i < 50$\deg)
despite the heavy line-of-sight absorption. Combining the X-ray data with infrared data
from \textit{WISE}, the results seen from \src\ further support the recent suggestion
that large X-ray and IR surveys may together be able to identify good lensed quasar
candidates in advance of detailed imaging studies.
\end{abstract}


\section{Introduction}

Strong gravitational lensing of distant quasars is a particularly powerful tool in
astrophysics and cosmology. Time delays between the multiple images of the quasar
produced by the foreground lens offer an opportunity to constrain $H_0$ (\eg\
\citealt{Chartas02, Suyu17}), and microlensing variations -- related to the motion of
small-scale structure in the lens -- can constrain both the emitting region sizes in the
quasars (\eg\ \citealt{Dai10, MacLeod15}) as well as the stellar populations and dark
matter distributions of the lenses (\eg\ \citealt{Bate11}). At a more basic level, the
overall magnification of the intrinsic flux means that high signal-to-noise (S/N) data can
more easily be collected for sources at cosmologically interesting distances. In the
X-ray band specifically, this has permitted studies of the innermost accretion flow and
black hole spin (\eg\ \citealt{Reis14nat, Reynolds14, Walton15lqso}), the outflows
launched by the accretion process (\eg\ \citealt{Chartas03}), and the properties of
AGN coronae (\citealt{Lanzuisi19}) for systems beyond the local universe.

2MASS\,J10422211+1641151, hereafter \src, is a rare example of a strongly lensed
quasar that is heavily reddened in the optical, implying strong obscuration towards the
central engine (\citealt{Glikman21}). Although a couple of hundred lensed quasars are
now known,\footnote{For example, the Gravitationally Lensed Quasar Database contains
$\sim$220 sources, see https://web1.ast.cam.ac.uk/ioa/research/lensedquasars} only a
handful appear to be intrinsically obscured systems despite the fact that most black hole
growth is now expected to occur during an obscured phase (see \citealt{Brandt15rev} for
a recent review); other examples include MG\,2016+112 (\citealt{Lawrence84}),
MG\,J0414+0534 (\citealt{Hewitt92}), MACS\,J212919.9-074218 (\citealt{Stern10}) and
MG\,1131+0456 (\citealt{Stern20}). \src\ is also a quadruply lensed system, which are
also particularly rare (\eg\ \citealt{Lemon19}). 

The quasar in \src, at $z = 2.517$, is radio-quiet (undetected in the FIRST survey), and
is lensed by an early-type foreground galaxy at $z_{\rm{lens}} = 0.599$, producing a
`cusp' configuration for the four quasar images with an Einstein ring of $\sim$0.9$''$
(\citealt{Glikman21}; broadly similar to \eg\ RX\,J0911$+$-0551, \citealt{Bade97}, and
RX\,J1131-1231, \citealt{Sluse03}).\footnote{\src\ is technically a short-axis cusp
(\citealt{Saha03}), and so is more similar to the configuration of RX\,J0911$+$-0551.}
Based on the \Ha\ and \Hb\ line widths, \cite{Matsuoka18} argue for the presence of a
very massive black hole with $\log[M_{\rm{BH}}/M_{\odot}] \sim 9.6-10$, leading them
to suggest the mass ratio of the black hole and its host galaxy is anomalously high
($M_{\rm{BH}}/M_{\rm{gal}} \sim 0.01-0.02$, based on the galaxy mass inferred from
modelling the optical--IR SED with appropriate AGN/host galaxy templates) when
compared to other obscured systems at similar redshifts (which typically have
$M_{\rm{BH}}/M_{\rm{gal}} \lesssim 0.004$; \citealt{Alexander08, Wu18}). \src\ is also
detected as a bright X-ray source. Based on a series of snapshot observations taken
with the \swiftng\ (hereafter \swift; \citealt{SWIFT}), \cite{Matsuoka18} report that the
X-ray spectrum implies fairly heavy obscuration, as expected based on its classification
as a red quasar, with an absorbing column density of $N_{\rm{H}} \sim 5 \times
10^{23}$\,\pcmsq. However, the low S/N of the \swift\ data prevent robust constraints
on the level of X-ray absorption when considering the latest physical absorption models
(\eg\ \citealt{borus}). Furthermore, \swift\ does not have the imaging capabilities to
resolve the different images of the quasar.

Here we report on new broadband X-ray observations of \src\ taken in 2019/20 with
\chandra\ (\citealt{CHANDRA}), \xmm\ (\citealt{XMM}) and \nustar\ (\citealt{NUSTAR}).
Throughout this work, we assume a standard $\Lambda$CDM concordance cosmology,
\ie \lambdaCDM.

\section{Observations and Data Reduction}
\label{sec_red}

\xmm, \nustar\ and \chandra\ performed a series of partially coordinated observations
(1, 5 and 2 exposures, respectively) of \src\ throughout late 2019 and early 2020. The
details of these observations are summarised in Table \ref{tab_obs}.

\subsection{Chandra}

Both of the \chandra\ observations were taken with the ACIS-S detector
(\citealt{CHANDRA_ACIS}). We reduced the data for each with \ciao\ v4.11
(\citealt{CHANDRA_CIAO}) and its associated calibration files. Cleaned event files were
generated as standard with the \chandrarepro\ script, with the \edser\ sub-pixel event
re-positioning algorithm enabled (\citealt{CHANDRA_EDSER}). These were then
re-binned to 1/8th of the ACIS pixel size before smoothing with a Gaussian (0.25$''$
FWHM) for visualisation. All of the spectra analysed here were extracted with the
\specextract\ script, which also generated the relevant instrumental response files; for
spectra of the individual quasar images, source regions of radius 0.5--0.7$''$ were
used (specifically we used radii of 0.7$''$ for images A and D, and a radius of 0.5$''$
for image C, taking care to ensure that these regions do not overlap), while for the
integrated emission from all of the images we used a larger region of radius 2$''$,
which encompasses all four quasar images. In all cases, the background was estimated
from larger regions (radius 20$''$) of blank sky on the same chip as \src.

\subsection{\textit{NuSTAR}}

All of the \nustar\ exposures were reduced with the \nustar\ Data Analysis Software
(\nustardas) v1.8.0. Cleaned event files were produced for both of the focal plane
modules (FPMA and FPMB) with \nupipeline, using instrumental calibration files from
\nustar\ \caldb\ v20190627 and the standard depth correction (which significantly
reduces the internal high-energy background). Passages through the South Atlantic
Anomaly were excluded using the following settings: {\small SAACALC}\,=\,3,
{\small SAAMODE}\,=\,None and {\small TENTACLE}\,=\,No. Source spectra were
extracted from the cleaned event files using circular regions of radius 50$''$ with
\nuproducts, which also generated the associated instrumental response files.
As with the \chandra\ data, background was estimated from larger regions of blank
sky on the same detector as \src. In order to maximise the exposure used, we
extracted both the standard `science' (mode 1) data and the `spacecraft science'
(mode 6) data (see \citealt{Walton16cyg} for a description of the latter). The mode 6
data provide $\sim$10\% of the total \nustar\ exposure times listed in Table
\ref{tab_obs}. Finally, given the moderate S/N of the data for the individual focal plane
modules, we combine the data for FPMA and FPMB together using \addascaspec. We
note that none of these \nustar\ exposures show evidence of abnormally low
temperatures for the optics bench related to issues with the apparent rip in its
insulation layers (\citealt{NUSTARmli}).

\begin{table}
  \caption{Details of the 2019/20 X-ray observations of \src.}
\begin{center}
\begin{tabular}{c c c c c c}
\hline
\hline
\\[-0.25cm]
Mission & OBSID & Start Date & Good & $F_{2-10}$\tmark[b] \\
\\[-0.35cm]
 & & & Exp.\tmark[a] & (obs frame) \\
\\[-0.25cm]
\hline
\hline
\\[-0.2cm]
\multicolumn{5}{c}{\textit{2019 Observations}} \\
\\[-0.25cm]
\xmm\ & 0852000101 & 2019-11-24 & 20/25 & $1.03 \pm 0.05$ \\
\\[-0.25cm]
\nustar\ & 60501032002 & 2019-11-24 & 55 & $1.16 \pm 0.05$ \\
\\[-0.1cm]
\multicolumn{5}{c}{\textit{2020 Observations}} \\
\\[-0.25cm]
\chandra\ & 22624 & 2020-01-27 & 22 & $1.12^{+0.07}_{-0.08}$ \\
\\[-0.25cm]
\chandra\ & 23135 & 2020-02-02 & 24 & $1.21^{+0.07}_{-0.08}$ \\
\\[-0.25cm]
\nustar\ & 60601001002 & 2020-02-02 & 7 & $1.3^{+0.1}_{-0.2}$ \\
\\[-0.25cm]
\nustar\ & 60601001004 & 2020-02-03 & 10 & $1.2 \pm 0.1$ \\
\\[-0.25cm]
\nustar\ & 60601001006 & 2020-02-06 & 22 & $1.17 \pm 0.07$ \\
\\[-0.25cm]
\nustar\ & 60601001008 & 2020-02-13 & 23 & $1.23 \pm 0.07$ \\
\\[-0.2cm]
\hline
\hline
\end{tabular}
\end{center}
$^{a}$ All exposures are given in ks (and rounded to the nearest whole value); \xmm\
exposures are quoted listed for the \epicpn/MOS detectors. \\
$^{b}$ Fluxes in the 2--10\,keV bandpass (observed frame, without correction for
absorption) in units of $10^{-12}$\,\ergpcmsqps\ for the individual observations,
integrated across all four quasar images (since \xmm\ and \nustar\ do not have the
spatial resolution to separate them).
\vspace*{0.3cm}
\label{tab_obs}
\end{table}

\subsection{\textit{XMM-Newton}}

The \xmm\ observation was also reduced following standard procedures, using the
\xmm\ Science Analysis System (\sas\ v18.0.0). Cleaned event files were produced
using \epchain\ and \emchain\ for the \epicpn\ and \epicmos\ detectors, respectively
(\citealt{XMM_PN, XMM_MOS}). All of the EPIC detectors were operated in full frame
mode. Source spectra were extracted from the cleaned eventfiles with \xmmselect\
using a circular region of radius 25$''$. As with both \chandra\ and \nustar, background
was estimated from larger regions of blank sky on the same chips as \src. The
background flaring was fairly severe for this observation, so we employed the method
outlined in \cite{Picon04} to determine the background level that maximises the S/N
for the source, and excluded periods that exceeded this threshold. This process was
performed independently for each of the \epicpn\ and the \epicmos\ detectors. During
the spectral extraction, we only considered single and double patterned events for
\epicpn\ ({\small PATTERN}\,$\leq$\,4) and single to quadruple patterned events for
\epicmos\ ({\small PATTERN}\,$\leq$\,12), as recommended. The instrumental
response files for each of the EPIC detectors were generated using \rmfgen\ and
\arfgen, and after performing the reduction separately for the two \epicmos\ units we
also combined these data using \addascaspec.

\section{Analysis}

\subsection{\textit{Chandra} Imaging}
\label{sec_img}

\begin{figure}
\begin{center}
\hspace*{-0.2cm}
\rotatebox{0}{
{\includegraphics[width=240pt]{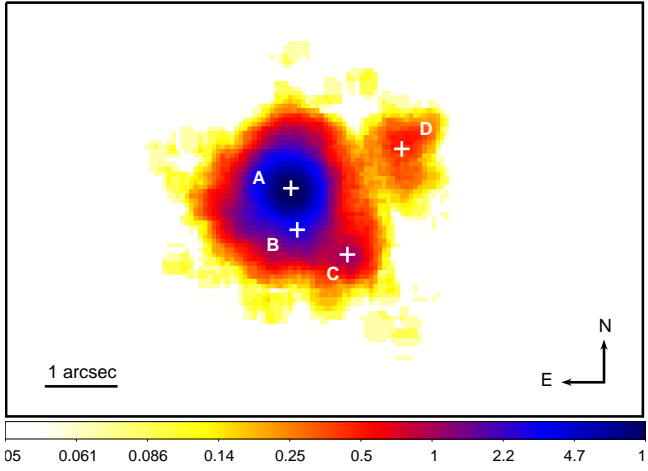}}
}
\end{center}
\vspace*{-0.3cm}
\caption{
The X-ray image of the quadruply lensed quasar \src\ obtained with \chandra\
(smoothed with a 0.375$''$ boxcar). The colour bar shows the scaling of the image
(in counts). Evidence for X-ray emission is seen from all four of the known images
of the quasar (positions indicated by white crosses, based on \citealt{Glikman21}).
Image A dominates the total flux, image B is just to the south (and is blended with
the emission from image A), image C is further to the south-west, and image D is
the furthest to the west.}
\label{fig_image}
\end{figure}

The combined \chandra\ image of \src\ from the two OBSIDs is shown in Figure
\ref{fig_image}. after ensuring that the two observations were aligned and
registered to a common coordinate system using \ciao. To do so, we produce X-ray
source lists (specifically excluding \src) for both OBSIDs individually using
\wavdetect, determine any relative offset between the two observations using
\wcsmatch, and correct the coordinate system of the second observation using
\wcsupdate. The transformation is determined by initially matching X-ray sources
within a 2$''$ radius, and then iteratively updating the astrometric solution to keep
only those that match within a radius of 0.5$''$ once the offsets are applied; note
that only translational corrections are considered here. The imaging data from each
of the individual observations are then combined using \reprojectobs. Based on
these combined data, quasar images A, C and D are clearly resolved by \chandra,
with image A dominating the total emission, as is also the case in the longer
wavelength data (\citealt{Glikman21}). Image B is heavily blended with the emission
from image A, but still contributes a detectable X-ray flux.

\subsubsection{Image Resolved Spectroscopy}

Given the visual inspection of the overall X-ray image, we initially attempt to extract
separate spectra from regions associated with images A, C and D. We see no evidence
for strong variability between the spectra from the two observations in any of these
images individually, and so for each we combined the data from the two OBSIDs
using \addascaspec. For images C and D, contamination by the wings of the
point-spread function (PSF) associated with image A is a potential concern. In order
to assess the level of this contamination we take two approaches; we investigate the
counts in a set of regions from a range of azimuthal angles to the north of image A
that have the same size and separation from image A as the regions centred on
images C and D, and we also use our more formal image modeling (see Section
\ref{sec_psf}) to predict the expected contamination from image A in the extraction
regions used for images C and D. Both approaches suggest that the expected level of
contamination is $\sim$15--20\% for the spectra extracted from both images C and
D, which, although not completely negligible, we consider to be a reasonably
manageable level. Any contamination from image B to the data from the image A
should also be negligible (see Section \ref{sec_psf}).

To test whether the spectra from the individual images are consistent, we model
these combined spectra with a phenomenological absorbed powerlaw model, fitting
the data over the 0.5--8\,keV energy range with \xspec\  v12.10.1s (\citealt{xspec}).
We include the Galactic absorption column ($N_{\rm{H, Gal}} = 2.18 \times
10^{20}$\,\pcmsq; \citealt{NH2016}), as well as a second absorber intrinsic to \src\ 
(\ie at $z = 2.517$); neutral absorption is modelled with \tbabs\ (\citealt{tbabs}). We
use the absorption cross-sections of \cite{Verner96}, and adopt the solar abundances
of \cite{Grevesse98} throughout this work for consistency with the \relxill\ model
(\citealt{relxill}), which is used in our later analysis.

Owing to the low S/N for images C and D (these spectra include 110 and 91 total
counts, respectively), at this stage in our analysis we rebin the individual spectra to 1
count per energy bin, and fit the data by reducing the Cash statistic (\citealt{cstat}).
We found that the values for the intrinsic column, $N_{\rm{H}}$, and the photon index,
$\Gamma$, are consistent within their 90\% uncertainties for all three images (note
that, unless stated otherwise, we quote 90\% errors as standard throughout this work).
We caution, however, that owing to the low S/N of the image C/D data and the
contaminating flux from image A, only large spectral variations between the quasar
images would be detectable, and that more subtle differences could still be present.
Nevertheless, we therefore also extracted the integrated spectrum combining all of
the images together, and fit this with the same model. For the integrated spectrum, we
have sufficient statistics to rebin the data to a minimum S/N of 5 (note that the S/N is
calculated accounting for the background and its uncertainties), and fit by minimising
\chisq. We find $N_{\rm{H}} = 2.7^{+0.6}_{-0.5} \times 10^{23}$\,\pcmsq\  and
$\Gamma = 1.10 \pm 0.15$. The fit is statistically very good, \chisq\ = 78 for 76
degrees of freedom (DoF), but despite the fairly large absorption column the photon
index is still unusually hard for a radio-quiet AGN (which typically have $\Gamma \sim
1.9$; \eg\ \citealt{Ricci17}). Allowing the absorption to be partially covering (using the
\partcov\ model within \xspec) only provides a marginal improvement in the fit
(\chisq/DoF = 73/75) and does not change the conclusion regarding the photon index
significantly; the best-fit column increases slightly to $N_{\rm{H}} = (3.9 \pm 1.1) \times
10^{23}$\,\pcmsq\ and has a covering factor of $C_{\rm{f}} = 95 \pm 3$\,\%, but we
still find $\Gamma = 1.26 \pm 0.20$.

Using the simpler model with a fully covering absorber and assuming the spectral
parameters quoted above are common to all the quasar images (although we stress
that the model is still a purely phenomenological description of the data at this stage),
we return to the data for the individual images to determine their X-ray flux ratios.
Similar to the nomenclature used in \cite{Glikman21}, we quote the image ratios
relative to image C. The flux ratios we find are A/C = $13.1^{+2.4}_{-1.9}$ and D/C =
$0.76^{+0.20}_{-0.16}$. These are consistent with the average flux ratios reported by
\cite{Glikman21} based on the longer wavelength \hst\ WFC3 imaging with the F125W
and F160W filters. We caution, though, that \cite{Glikman21} find evidence that the
optical/IR flux ratios may be mildly variable (by up to $\sim$50\%) on timescales
comparable to the separation of the X-ray observations considered here (although
they also note that this could be at least in part related to modelling issues related to
the diffraction spikes from image A, which intercept the other quasar images during
their second \hst\ epoch); similar variability would certainly be permitted by these
results, particularly thanks to the low S/N for images C and D.

\subsubsection{Image Modelling}
\label{sec_psf}

In addition to considering these image-resolved spectra, we also perform a more direct
imaging analysis of the \chandra\ data to determine the image flux ratios more robustly.
For each observation a model of the \chandra\ PSF is simulated at the position of \src\
with \textsc{MARX v5.5.0} using the average source spectrum assuming the
fully-covering absorber described above and the observation-specific attitude file as
input. We then perform two-dimensional profile fits to the \chandra\ images with the
\textsc{sherpa} modelling and fitting package, using the Cash statistic and the
Nelder-Mead algorithm to determine the best fit and evaluate parameter uncertainties.
The image PSF is convolved with a model consisting of four Gaussians. The relative
positions of these Gaussians are fixed based on the relative positions of the quasar
images based on the \hst\ astrometry (\citealt{Glikman21}), but for each observation
a global positional offset is allowed to account for any differences in the overall
astrometric solutions between the \chandra\ and \hst\ observations. A constant was
added to the source model to account for the small background contribution in the
\chandra\ image. This model is then fit to the data from the respective \chandra\
exposures. The fits were carried out in 80$\times$80 pixel region centred on the QSO
position. The Gaussian widths are fixed between all lensed images and between both
observations. The free parameters are the normalisations of the Gaussian components
and the background constant, the global width of the Gaussian components,
and the positional offsets. In both observations, we find that the \chandra\ data prefer
a small astrometric shift of $\sim$0.04$''$ relative to \hst.

The flux ratios are initially calculated for the two \chandra\ observations separately, and
the results obtained are presented in Table \ref{tab_fluxrat}. Ultimately, though, the
results for the two epochs are consistent at the 90\% level, and so we combine the two
epochs together to determine our final constraints. The results from this analysis for the
A/C and D/C ratios are consistent with those found from the spectra extracted for these
images. As noted above, we also use the above model to try and predict the level of
contamination from the wings of the PSF From image A in the extraction regions used
for images C and D by turning the emission from image A and comparing the predicted
fluxes in image C/D regions to that of the best-fit model; we find the predicted level of
contamination to be $\sim$15--20\%. Indeed, the consistency of the flux ratios between
this imaging-based analysis (which properly considers the blending of the PSFs) and the
analysis of the spectra from the different image regions above (which does not) further
suggests that contamination of images C and D by the wings of the PSF of image A is not
too severe, and thus should not have a major impact on the comparison of the spectra
from these images discussed above.

We also note that the X-ray flux ratios from this analysis are again very similar to the
average ratios reported by \cite{Glikman21} based on the \hst\ WFC3 imaging
(although we caution again that these authors find the flux ratios may be mildly variable
at optical/IR wavelengths). We therefore conclude that the same flux anomaly seen at
longer wavelengths (\ie the fact that image A is by far the brightest\footnote{For an
ideal cusp configuration image B should be the brightest, while images A and C should
have similar fluxes (\citealt{Keeton03}).}) is also present in the X-ray band.

\begin{table}
  \caption{X-ray flux ratios for the four lensed images of \src}
\begin{center}
\begin{tabular}{c c c c c c}
\hline
\hline
\\[-0.25cm]
Epoch & A/C & B/C & D/C \\
\\[-0.25cm]
\hline
\hline
\\[-0.2cm]
1 & $9.6^{+3.0}_{-3.1}$ & $1.6^{+0.6}_{-0.7}$ & $0.4 \pm 0.2$ \\
\\[-0.25cm]
2 & $17.0^{+5.8}_{-6.2}$ & $2.0 \pm 1.0$ & $0.7 \pm 0.3$ \\
\\[-0.25cm]
Combined & $13.3^{+3.2}_{-3.5}$ & $1.8 \pm 0.6$ & $0.6 \pm 0.2$ \\
\\[-0.2cm]
\hline
\hline
\end{tabular}
\end{center}
\label{tab_fluxrat}
\end{table}

\subsection{Broadband Spectroscopy}
\label{sec_spec}

\begin{figure}
\begin{center}
\hspace*{-0.35cm}
\rotatebox{0}{
{\includegraphics[width=235pt]{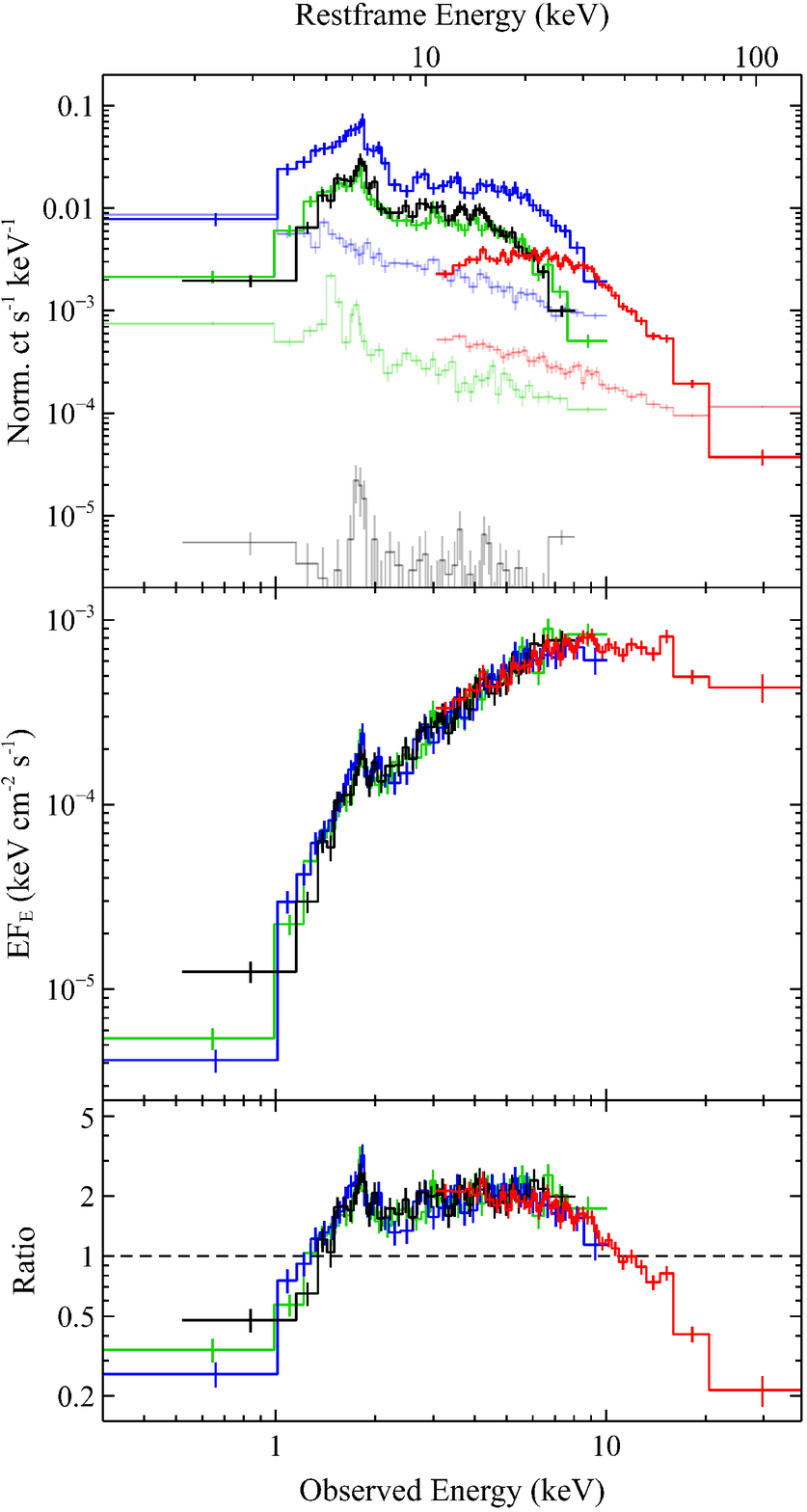}}
}
\end{center}
\vspace*{-0.3cm}
\caption{
The broadband X-ray spectrum of \src. The top panel shows the raw count spectra,
with data from the \chandra\ ACIS-S detector, the \nustar\ FPMs, and the \xmm\
\epicpn\ and \epicmos\ detectors are shown in black, red, blue and green, respectively.
The associated background levels for each detector are shown in the lighter shading
for each colour. \src\ is detected up to $\sim$40\,keV in the observed frame by \nustar,
corresponding to $\sim$140\,keV in the rest-frame of \src. The middle panel shows
these data unfolded through a model that is constant with energy, and the bottom
panel shows the data as a ratio to a fit (to the full band) with a phenomenological
powerlaw continuum, modified only by Galactic absorption ($\Gamma \sim 0.8$,
$N_{\rm{H,Gal}} =  2.18 \times 10^{20}$\,\pcmsq). These panels show clear iron
emission, and highlight the strong curvature in the continuum emission at higher
energies (peaking at $\sim$7\,keV in the observed frame). The data in all panels have
been further rebinned for visual clarity.}
\label{fig_spec_all}
\end{figure}

Having found there are no discernible differences between the spectra extracted from
the regions dominated by the different quasar images, we now turn to broadband
spectroscopy, incorporating the \xmm\ and \nustar\ data. Neither of these missions are
able to resolve the separate images of the quasar, so we compare them against the
integrated \chandra\ data from all of the images. We see no significant variability
between the \chandra\ and \xmm\ datasets. Similarly, we see no evidence for variability
between any of the \nustar\ observations (Table \ref{tab_obs}), so we combine the data
from all of these into one spectrum (again using \addascaspec). As with the integrated
\chandra\ data, we rebin the \xmm\ and integrated \nustar\ data to have a S/N of 5 per
energy bin in the same manner. We show all of these datasets (\chandra, \xmm\ and
\nustar) in Figure \ref{fig_spec_all}. \src\ is detected up to $\sim$40\,keV in the \nustar\
data (corresponding to a rest-frame energy of $\sim$140\,keV for \src). Since the
different datasets are all consistent with one another, in the following sections we
explore models for the broadband spectrum by modelling them simultaneously. As is
standard, during our analysis we allow multiplicative constants to float between them
to account for cross-calibration issues, fixing the constant for the \chandra\ data to
unity. The rest of these constants are always within $\sim$5\% of unity, as expected
based on detailed cross-calibration studies comparing the facilities used here
(\citealt{NUSTARcal}).

\begin{figure}
\begin{center}
\hspace*{-0.35cm}
\rotatebox{0}{
{\includegraphics[width=235pt]{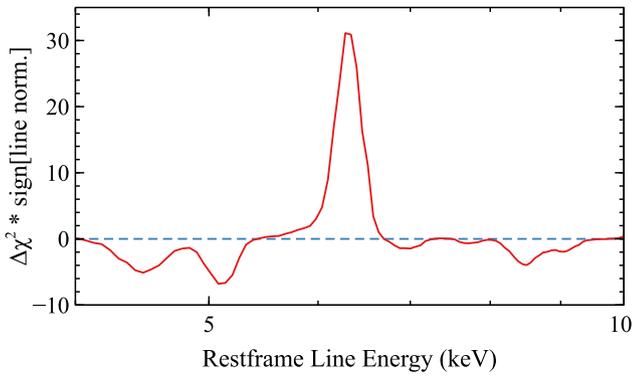}}
}
\end{center}
\vspace*{-0.3cm}
\caption{
Results from a Gaussian line scan applied to the combined \xmm\ and \chandra\ data
over the 4--10\,keV band, assuming a simple absorbed powerlaw continuum. The line is
assumed to be unresolved and its energy is varied in increments of 60\,eV, with the
improvement in \chisq\ noted for each step. Positive values of $\Delta\chi^{2}$ indicate
emission features, and negative values indicate absorption. The only feature of any note
is the Fe K emission line at $\sim$6.4\,keV.}
\label{fig_linescan}
\end{figure}

As indicated by the phenomenological fits to the \chandra\ data, the source spectrum
is clearly very hard, as expected for an absorbed source, but also exhibits strong
curvature at higher energies, peaking at $\sim$7\,keV (observed frame). This
corresponds to $\sim$25\,keV in the quasar rest-frame, and likely indicates a significant
contribution from Compton reflection (\ie the `Compton hump' that is characteristic for
reprocessing by Compton-thick material, \eg\ \citealt{George91}). There is also evidence
for narrow iron emission at $\sim$1.8\,keV (rest-frame energy of $\sim$6.4\,keV). Indeed,
fitting the rest-frame 2--10\,keV energy range with the absorbed powerlaw model
described above, but here utilizing the full \xmm\ and \chandra\ data (the relevant
energies are outside of the \nustar\ bandpass), the addition of a narrow (zero-width)
Gaussian emission line at 6.4\,keV improves the fit by $\Delta\chi^{2}$ = 26 for one
additional free parameter (the continuum parameters are consistent with those reported
above for just the \chandra\ data). Each of the ACIS, EPIC-pn and EPIC-MOS datasets
contribute roughly evenly to this improvement. We find a rest-frame equivalent width of
$EW = 220 \pm 80$\,eV (and similar continuum parameters to those quoted above). This
is somewhat intermediate to the values typically seen from unobscured type-1 AGN ($EW
\sim 100$\,eV; \eg\ \citealt{Bianchi07}) and the most obscured, Compton-thick AGN
($EW \sim 1$\,keV; \eg\ \citealt{Boorman18}). We also show in Figure \ref{fig_linescan}
the results of a broader emission/absorption line search, obtained by scanning a narrow
Gaussian line across the 4--10\,keV bandpass in the restframe of \src, allowing for the
line to be in either emission or absorption. Aside from the strong statistical improvement
provided by the Fe K emission, the available data do not show any compelling evidence
for any other line features.

Fitting now the full observed energy range (combining the 0.3--10\,keV data from \xmm,
the 0.5--8\,keV data from \chandra\ and the 3--40\,keV data from \nustar\ to give a total
coverage of 0.3--40\,keV in the observed frame, corresponding to 1--140\,keV in the
rest-frame of \src) and including an exponential high-energy cutoff in the powerlaw
continuum, we find this phenomenological model provides a reasonably good description
of the data, with \chisq/DoF = 432/383. However, we still find the same issues with the
continuum parameters as when considering the more limited rest-frame 2--10\,keV
energy range. Although the absorption column remains similar, $N_{\rm{H}} = (2.5 \pm
0.4) \times 10^{23}$\,\pcmsq, the best-fit photon index has actually become even harder,
$\Gamma = 0.44 \pm 0.13$. The high-energy cutoff is \ecut\ = $23^{+3}_{-2}$\,keV
(evaluated in the rest-frame of \src), which is also abnormally low; AGN typically have
\ecut\ $\gtrsim$ 50--100\,keV (\eg\ \citealt{Fabian15}, \citealt{Tortosa18},
\citealt{Balokovic20}, although there are rare exceptions, \eg\ \citealt{Kara17}). This again
suggests that the curvature seen at high-energies is at least partially set by a significant
contribution from Compton reflection. As before, these results do not change significantly
if we allow the absorption to be partially covering. They also do not change significantly if
we allow the absorption to be partially ionised instead (using the \xstar\ photoionisation
code\footnote{Specifically we use the grid of absorption models discussed in
\cite{Walton20iras}, which are broadly relevant for AGN.}; \citealt{xstar}).

\subsubsection{Torus Modelling}
\label{sec_borus}

Given that the simple absorbed powerlaw models considered above for the \chandra\ 
data all imply unusually hard photon indices, and the broadband data also imply the
presence of reprocessing, we now test more complex models that include both
processes. We start by exploring whether the broadband data can be explained by
absorption and reprocessing in an obscuring torus, as invoked in the classic unified
model for AGN (\eg\ \citealt{AGNunimod}). In particular, we utilize the \borus\ model
(\citealt{borus}), one of the most recent additions to the family of X-ray torus models
(and an update of the \torus\ model presented by \citealt{torus}). This model
self-consistently computes the absorption and reprocessing assuming the obscuring
medium has a uniform spherical geometry with conical polar cut-outs, is neutral, and is
illuminated internally by a central continuum flux. In general, its key free parameters are
the radial column density ($N_{\rm{H,tor}}$), iron abundance ($A_{\rm{Fe}}$) and
opening angle of the torus ($\theta_{\rm{tor}}$), the viewing angle ($i$) and the
primary continuum parameters (see below).

We initially test a model in which we are viewing the source through the torus, such that
the column along our line-of-sight is the same as that of the torus (\ie $N_{\rm{H,los}} = 
N_{\rm{H,tor}}$). We again use \tbabs\ for the line-of-sight photoelectric absorption,
combined with \cabs\ to account for the scattering losses in the absorber, both of which
are applied to the intrinsic AGN continuum. In addition, we also include both the
reprocessed emission from \borus\ and a fraction of the intrinsic continuum that is
scattered around the absorber. These last two components are not subject to absorption
by the main absorber, and are only subject to the Galactic absorption column (again fixed
at $N_{\rm{H, Gal}} = 2.18 \times 10^{20}$\,\pcmsq). This is a standard \borus\ setup for
studying heavily obscured AGN (\eg\ \citealt{borus}). In \xspec\ parlance, the model
expression is \tbabs$_{\rm{Gal}}$ $\times$ $($\borus\ $+$ {\small CONT}$_{\rm{scat}}$
$+$ $($\tbabs$_{\rm{los}}$ $\times$ \cabs\ $\times$ {\small CONT}$_{\rm{int}}))$,
where {\small CONT} indicates the AGN continuum model. To ensure we are looking
through the obscurer, we first employ a version of \borus\ that assumes the absorber
covers the full $4\pi$ solid angle (\ie has a spherical geometry). Formally, we use the
{\borus}11 model, which uses the \nthcomp\ thermal Comptonisation model for the
intrinsinc AGN continuum (parameterised by $\Gamma$ and the electron temperature,
\kte; \citealt{nthcomp1, nthcomp2}), again evaluated in the rest-frame of \src. Note that
because the geometry is spherical here, the viewing angle is not a free parameter (the 
illuminating continuum is also assumed to be isotropic in \borus). All of the relevant
model parameters are linked between the intrinsic AGN continuum, the \borus\ model,
and the scattered AGN continuum (\ie $\Gamma$, \kte, continuum
normalisations)\footnote{Note, however, that while \borus\ is normalised in the
rest-frame of the source, \nthcomp\ is normalised in the observed frame. In order to
meaningfully link their normalisations for sources with non-negligible redshifts, it is
necessary to set the redshift parameter in the \nthcomp\ components to zero, and
redshift them separately using \zashift\ components in \xspec.}. For the latter, the
scattered fraction is computed via a further multiplicative constant. The iron
abundances of the absorption and \borus\ components are also linked, after scaling the
\borus\ iron abundance to that of \cite{Grevesse98} for self-consistency (\borus\ is
formally calculated assuming the solar abundances of \citealt{Anders89}).

\begin{table}
  \caption{Results obtained with the torus models for \src.}
\begin{center}
\begin{tabular}{c c c c}
\hline
\hline
\\[-0.2cm]
Parameter & Units & \multicolumn{2}{c}{Model Value} \\
\\[-0.35cm]
& & {\borus}11 & {\borus}12 \\
\\[-0.25cm]
\hline
\hline
\\[-0.15cm]
$N_{\rm{H,los}}$ & [$10^{23}$\,\pcmsq] & $3.0 \pm 0.3$ & $3.8^{+0.5}_{-0.3}$ \\
\\[-0.3cm]
$A_{\rm{Fe}}$ & [solar] & $1.1^{+0.3}_{-0.2} $ & $0.55^{+0.17}_{-0.10}$ \\
\\[-0.3cm]
$N_{\rm{H,tor}}$ & [$10^{23}$\,\pcmsq] & = $N_{\rm{H,los}}$ & $>31$ \\
\\[-0.3cm]
$\Omega_{\rm{tor}}/4\pi$ & & 1.0\tmark[a] & $>0.61$\tmark[b] \\
\\[-0.3cm]
$i$ & [\deg] & -- & 20\tmark[a] \\
\\[-0.3cm]
$\Gamma$ & & $<1.41$\tmark[c] & $1.72^{+0.04}_{-0.12}$ \\
\\[-0.3cm]
\kte\ & [keV] & $11.3^{+0.5}_{-0.4}$ & $26^{+7}_{-9}$ \\
\\[-0.3cm]
Norm & [$10^{-4}$] & $7.0^{+0.4}_{-0.3}$ & $16^{+5}_{-3}$ \\
\\[-0.3cm]
$f_{\rm{scat}}$ & [\%] & $4.4 \pm 1.2$ & $1.6 \pm 0.7$ \\
\\[-0.2cm]
\hline
\\[-0.2cm]
\chisq/DoF & & 381/382 & 359/380 \\
\\[-0.25cm]
\hline
\hline
\end{tabular}
\label{tab_borus}
\end{center}
\flushleft
$^{a}$ Indicates the parameter is fixed at this value. \\
$^{b}$ Note that $\Omega_{\rm{tor}}/4\pi$ is limited to $< 0.9$ in this fit. \\
$^{c}$ \borus\ is limited to photon indices $\Gamma > 1.4$. \\
\end{table}

This model formally fits the data very well, with \chisq\ = 381 for 382 DoF; the full results
are presented in Table \ref{tab_borus}. The column density of $N_{\rm{H,tor}} \sim 3
\times 10^{23}$\,\pcmsq\ is similar to that found considering just the \chandra\ data, and
the scattered continuum fraction is $f_{\rm{scat}} = 4.4 \pm 1.2$\%, broadly similar to
the values typically seen in local AGN (which are also at the level of a few per cent, \eg\
\citealt{Winter09, Eguchi09, Walton18, Walton19ufo, Kammoun20}). However, the
continuum parameters are still abnormal for a radio-quiet AGN. The photon index of
$\Gamma < 1.41$ (note that \borus\ is only calculated for $\Gamma > 1.4$) is still much
harder than would be expected for such sources. Furthermore, in order to reproduce
the strong high-energy curvature with such a hard continuum at lower energies, the
electron temperature of \kte\ = $11.3^{+0.5}_{-0.4}$\,keV is also lower than usually seen
in local AGN (as noted above, typically \ecut\ $\gtrsim$  50--100\,keV, corresponding
to \kte\ $\gtrsim$ 30\,keV).\footnote{Typical conversion factors are \ecut\ $\sim$
2--3\,\kte, see \eg\ \cite{Petrucci01}.} This is in large part because the torus is required
to be Compton-thin in this scenario, and so does not provide any notable reprocessing
at high-energies. If we force the column to be Compton-thick here, \ie\ $N_{\rm{H}} >
1.5 \times 10^{24}$\,\pcmsq, the fit degrades to \chisq/DoF = 1321/382. Although we
show the results for a spherical obscurer, we stress that these conclusions do not
change if we instead use a version of \borus\ with a variable covering factor (\ie
{\borus}12, which again uses the \nthcomp\ model for the intrinsic continuum;
\citealt{borus12}), as long as the requirement that the column density of the torus is the
same as the line-of-sight column density is retained.

\begin{figure}
\begin{center}
\hspace*{-0.35cm}
\rotatebox{0}{
{\includegraphics[width=235pt]{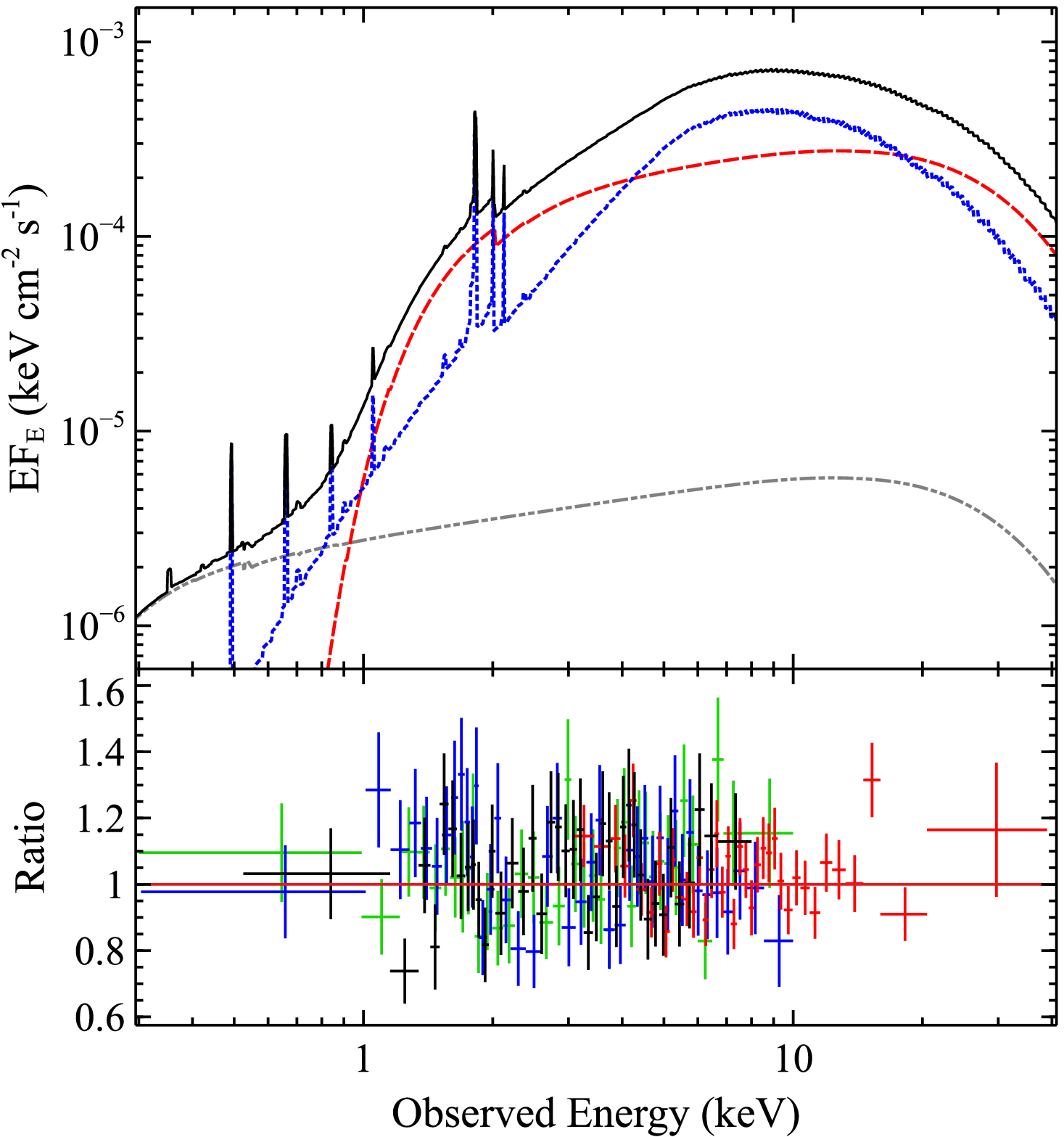}}
}
\end{center}
\vspace*{-0.3cm}
\caption{
The torus model fit to the \chandra+\nustar+\xmm\ data for \src, using the {\borus}12
model (see Table \ref{tab_borus} for the best-fit parameters). The top panel shows the
relative contribution of the different model components: the total model is shown in
solid black, the primary powerlaw emission in dashed red, the reprocessed emission
from the torus in dotted blue, and the scattered continuum in dash-dot-dot grey. The
bottom panel shows the data/model ratios for the broadband dataset; here the different
colours have the same meanings as in Figure \ref{fig_spec_all}.}
\label{fig_borus}
\end{figure}

We therefore test the alternative assumption, that we are not viewing the central source
through the bulk of the reprocessing torus, which instead primarily lies out of our
line-of-sight (\ie $N_{\rm{H,los}} < N_{\rm{H,tor}}$), continuing with the use of the
{\borus}12 model. In order to ensure we do not view the source through the torus, we fix
the viewing angle for \borus\ at $i = 20$\deg\ (roughly the minimum allowed by the
model), and limit its covering factor such that the reprocessing torus cannot impinge on
this line-of-sight (\ie we set a limit of $\Omega_{\rm{tor}}/4\pi < 0.9$); the rest of the
model setup is the same as described previously. This approach also gives a good fit to
the data, with \chisq/DoF = 359/380, and as before the full results are given in Table
\ref{tab_borus}. The line-of-sight column density is similar to that inferred previously,
$N_{\rm{H,los}} = 3.8^{+0.5}_{-0.3} \times 10^{23}$\,\pcmsq, but, as expected, the
column density inferred for the main torus is much larger, $N_{\rm{H,tor}} > 3.1 \times
10^{24}$\,\pcmsq. The covering factor of this torus must be fairly large,
$\Omega_{\rm{tor}}/4\pi > 0.61$, in order to produce the strong high-energy curvature,
despite the fact that we cannot be viewing the source through an absorber with such a
large column. The iron abundance in this case is mildly sub-solar, as the equivalent width
of the narrow iron emission is not particularly large. However, thanks to the strong 
contribution from the reprocessed continuum, the continuum parameters for this model
($\Gamma$, \kte) are significantly closer to those expected from a radio-quiet AGN
(although the electron temperature is still a little on the low side). We show this model and
the corresponding data/model ratio in Figure \ref{fig_borus}.

Given that \cite{Glikman21} report the presence of a broad absorption line (BAL) in the
optical spectrum, specifically on the blue wing of the broad \mgii\ emission line, we again
test whether the X-ray absorption could be associated with partially ionised material in an
outflow now that we are considering the full broadband spectrum. Replacing the neutral
absorber with a photoionised absorption model, using the same \xstar\ as used previously,
does not provide any statistical improvement over the neutral absorber, and the key
parameters for the AGN emission remain essentially identical to those reported in Table 
\ref{tab_borus}.

\begin{table}
  \caption{Results obtained with the relativistic disc reflection model for \src.}
\begin{center}
\begin{tabular}{c c c c}
\hline
\hline
\\[-0.2cm]
Parameter & Unit & Model Value \\
\\[-0.25cm]
\hline
\hline
\\[-0.15cm]
$N_{\rm{H,los}}$ & [$10^{23}$\,\pcmsq] & $3.2^{+0.5}_{-0.4}$ \\
\\[-0.3cm]
$A_{\rm{Fe}}$ & [solar] & $0.9 \pm 0.3$ \\
\\[-0.3cm]
$\Gamma$ & & $1.61^{+0.19}_{-0.07}$ \\
\\[-0.3cm]
\kte\ & [keV] & $31^{+126}_{-8}$ \\
\\[-0.3cm]
$h$ & [\rg] & 5\tmark[a] \\
\\[-0.3cm]
$a^*$ & & 0.7\tmark[a] \\
\\[-0.3cm]
$i$ & [\deg] & $< 36$ \\
\\[-0.3cm]
$\log\xi$ & $\log$[\ergcmps] & 0\tmark[a] \\
\\[-0.3cm]
$R_{\rm{frac}}$ & & $1.7^{+2.9}_{-0.5}$ \\
\\[-0.3cm]
Norm & [$10^{-5}$] & $2.7^{+0.4}_{-0.8}$ \\
\\[-0.3cm]
$EW_{\rm{Fe K}}$\tmark[b] & [eV] & $170^{+100}_{-90}$ \\
\\[-0.3cm]
$f_{\rm{scat}}$ & [\%] & $2.9^{+0.9}_{-0.8}$ \\
\\[-0.2cm]
\hline
\\[-0.2cm]
\chisq/DoF & & 352/379 \\
\\[-0.25cm]
\hline
\hline
\end{tabular}
\label{tab_relxill}
\end{center}
\flushleft
$^{a}$ Indicates the parameter is fixed at this value. \\
$^{b}$ This is for the narrow core of the iron line. \\
\end{table}

\subsubsection{Disc Reflection Modelling}
\label{sec_relxill}

We also test the scenario in which the majority of the reprocessing occurs in the
accretion disc instead of a Compton-thick torus. In this case, the rapid orbital motion in
the disc and the strong gravity close to the black hole broaden and skew the narrow,
rest-frame emission lines into a characteristic `diskline' profile (\eg\ \citealt{Fabian89,
relconv}). Evidence for reflection from the accretion disc has been seen in other, less
obscured lensed quasars in the form of such relativistically broadened iron emission
lines (\eg\ \citealt{Reis14nat, Reynolds14, Walton15lqso}).


For this analysis we use the \relxilllpcp\ model (v1.3.3; \citealt{relxill}), which
self-consistently calculates the expected reflection spectrum assuming a lamppost
geometry (characterised by the height of the X-ray source above the disc, $h$) and
assumes the \nthcomp\ model for the primary X-ray continuum (which is also included
in the model.\footnote{Note that here \kte\ is evaluated in the rest-frame of the quasar
X-ray source, accounting for both the cosmological redshift of \src\ and the additional
gravitational redshift associated with the assumed accretion geometry.} The other key
model parameters are the spin of the black hole, $a^*$, the reflection fraction,
$R_{\rm{frac}}$ (which sets the relative contribution of the reflected emission; see
\citealt{relxill_norm} for the definition of \Rfrac\ used in \relxilllpcp), and the inner and
outer radii, inclination, iron abundance and ionisation parameter, $\xi$, of the disc. As
is standard, the ionisation parameter is defined as $\xi = L_{\rm{ion}} / n  R^{2}$,
where $L_{\rm{ion}}$ is the ionising luminosity (integrated between 0.1--1000\,keV in
the \relxill\ models), $n$ is the density of the material, and $R$ is the distance to the
ionising source.

\begin{figure}
\begin{center}
\hspace*{-0.35cm}
\rotatebox{0}{
{\includegraphics[width=235pt]{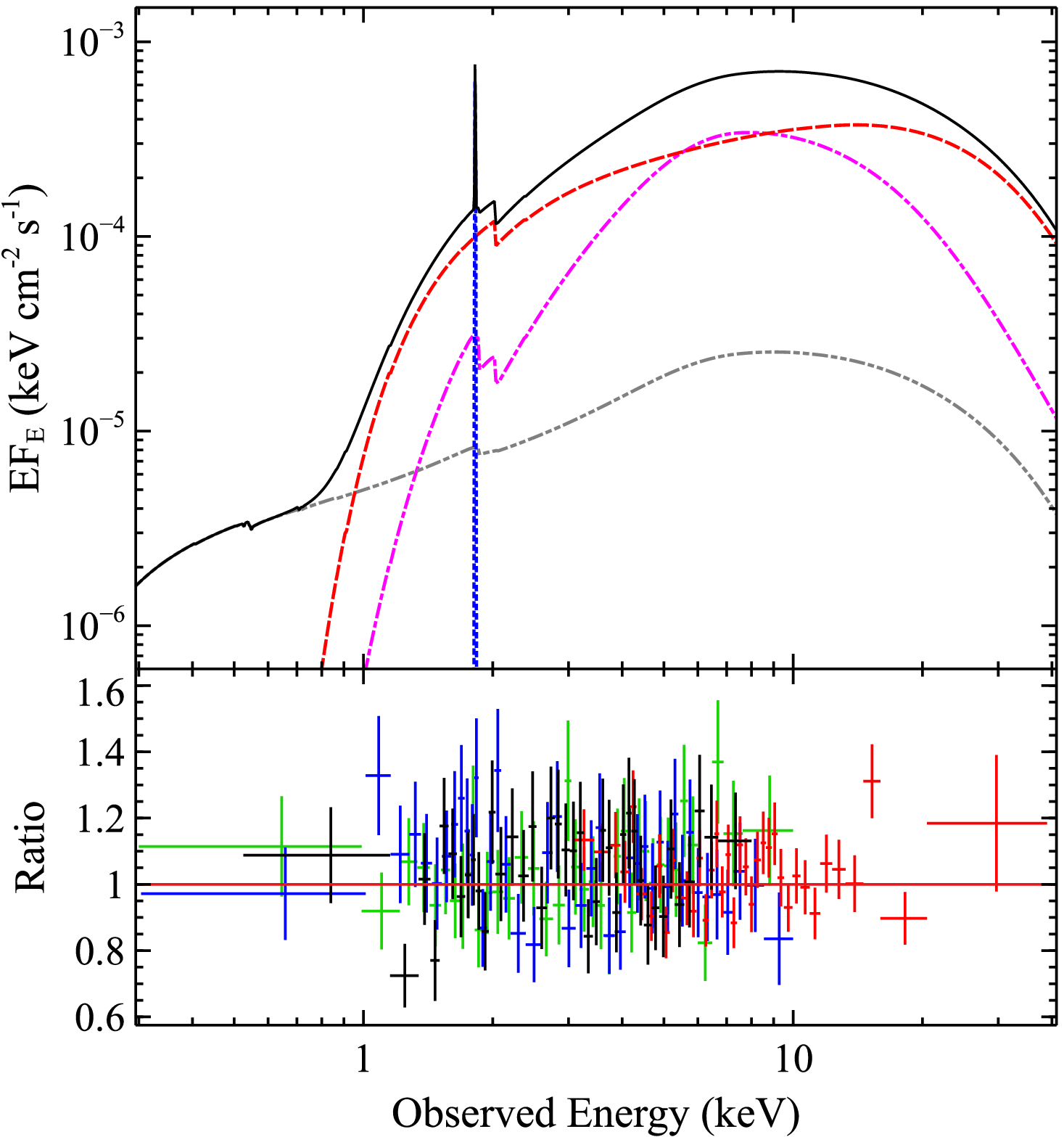}}
}
\end{center}
\vspace*{-0.3cm}
\caption{
The disc reflection model fit to the \chandra+\nustar+\xmm\ data for \src\ (see Table
\ref{tab_relxill} for the best-fit parameters). As in Figure \ref{fig_borus}, the top panel
shows the relative contribution of the different model components: the total model is
shown in solid black, the primary \nthcomp\ continuum in dashed red and the disc
reflection in dash-dot magenta (which together form the \relxilllpcp\ model), the narrow
Fe K line in dotted blue, and the scattered continuum in dash-dot-dot grey. The bottom
panel again shows the data/model ratios, and the colours have the same meanings as in
Figure \ref{fig_spec_all}.}
\label{fig_relxill}
\end{figure}

As before, we include neutral absorption from our own Galaxy, and allow for neutral
absorption in the rest-frame of the source, as well as some scattered fraction of the
intrinsic emission that leaks around the rest-frame absorber. We also include a narrow
core to the iron emission, which still likely arises from reprocessing by distant material
(and so is not subject to the relativistic effects modelled by \relxilllpcp); here we treat
this as a Gaussian emission line for simplicity. Overall, the model setup is similar to that
outlined in Section \ref{sec_borus}, where {\small CONT} equates to the \relxilllpcp\
model here, and \borus\ is replaced by the Gaussian component. During our analysis,
we assume that the inner accretion disc reaches the innermost stable circular orbit
(ISCO), and fix the outer disk to 1000\,\rg\ (the maximum allowed by the model; \rg\ =
$GM_{\rm{BH}}/c^2$ is the gravitational radius). As before, we also link the iron
abundance between the rest-frame absorber and the \relxilllpcp\ components.
Furthermore, the data do not have sufficient S/N to constrain the other key parameters
that control the precise form of the relativistic blurring ($a^*$ and $h$), so we fix these
to $a^* = 0.7$ and $h = 5$\,\rg, respectively; the former is the spin implied by the
average radiative efficiency of $\eta = 0.1$ inferred for quasars (\citealt{Soltan82}), and
the latter is motivated by constraints on X-ray emitting regions from micro-lensing
studies of other, less obscured lensed quasars (\eg\ \citealt{Dai10, MacLeod15}). If we
allow the spin to vary, we find it to be completely unconstrained by the current data. We
also find the ionisation of the disc is not well constrained, so for simplicity we assume
that the disk is close to being neutral (\ie $\log[\xi/(\rm{erg}~\rm{cm}~\rm{s}^{-1})] = 0$).

This model also fits the data very well, with \chisq/DoF = 353/379; the results are
presented in Table \ref{tab_relxill}, and the fit is shown in Figure \ref{fig_relxill}. The
line-of-sight column density is similar to that found previously, $N_{\rm{H,los}} =
3.2^{+0.5}_{-0.4} \times 10^{23}$\,\pcmsq, and the primary continuum parameters
($\Gamma$, \kte) are again consistent with expectation for a radio-quiet AGN. The
reflection fraction of $R_{\rm{frac}} = 1.7^{+2.9}_{-0.6}$ is consistent with both
standard values for local, unobscured AGN ($R_{\rm{frac}} \sim 1-1.5$, \eg\
\citealt{Walton13spin}, corresponding to the rough expectation for standard illumination
of a thin accretion disc, \eg\ \citealt{relxill_norm}) or a reflection-dominated scenario
($R_{\rm{frac}} > 2$). In the context of disc reflection, the latter scenario is possible if
the corona is particularly compact, such that the primary X-ray continuum emission
experiences strong gravitational lightbending (\eg\ \citealt{lightbending}), and examples
of reflection-dominated states have been seen among local AGN (\eg\
\citealt{Parker14mrk, Walton20iras}). We also note that the best-fit iron abundance in
this scenario is close to the solar value. Interestingly, the inclination of the accretion
disc is inferred to be low with this model, $i < 36$\deg, \ie we would be viewing the disc
close to face-on, despite the fairly heavy line-of-sight obscuration.

Finally, we note that we again tried replacing the neutral absorber in the disc reflection
model with the \xstar\ photoionisation model used previously, but as with the torus model
this did not result in any statistical improvement to the fit, and all of the key parameters
for the AGN emission remain the same as reported in Table \ref{tab_relxill}.

\section{Discussion \& Conclusions}
\label{sec_dis}

\begin{figure}
\begin{center}
\hspace*{-0.35cm}
\rotatebox{0}{
{\includegraphics[width=247.5pt]{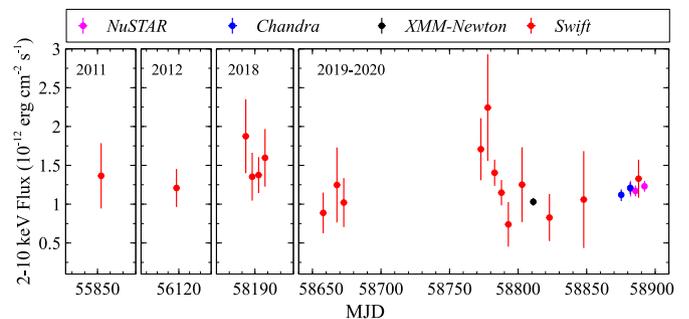}}
}
\end{center}
\vspace*{-0.45cm}
\caption{
The long-term X-ray lightcurve of \src\ in the 2--10\,keV band (observed frame). The
\xmm, \chandra\ and \nustar\ fluxes are those reported in Table \ref{tab_obs}, while
the \swift\ data were originally generated using the online pipeline (\citealt{Evans09})
with 5\,d time bins, and count rates then converted to flux based on the simple
absorbed powerlaw fit to the combined 2--10\,keV \xmm, \chandra\ and \nustar\ data
presented in Section \ref{sec_spec}.}
\label{fig_lc}
\end{figure}

We have presented the first sensitive X-ray view of the reddened, strongly lensed quasar
\src\ ($z = 2.517$), combining data from the \chandra, \xmm\ and \nustar\ observatories.
This is a quadruply lensed system, based on optical/IR imaging (\citealt{Glikman21}), and
the \chandra\ data show clear X-ray emission from each of the four quasar images (see
Figure \ref{fig_image}). The flux ratios are consistent with those seen at longer
wavelengths, and, within the limitations of the available data, we do not see any evidence
for large differences in their X-ray spectra (which can occur if some of the intervening
absorption occurs in the lensing galaxy, \eg\ the cases of B1152+199 and
GraL\,J234330.6+043557.9; \citealt{Toft00, Dai09, KroneMartins20}). This is consistent
with the absorption being intrinsic to the background quasar, as concluded by
\cite{Glikman21} and assumed throughout this work. The X-ray flux ratios show that the
flux anomaly seen at longer wavelengths, in which image A dominates the total flux, is
also clearly seen in the X-ray data. \cite{Glikman21} suggest that this may be due to
substructure (as predicted by $\Lambda$CDM) surrounding the main lensing galaxy
(\eg\ \citealt{Mao98}), as opposed to the microlensing alternative, as they argue that the
flux anomaly is still seen at longer wavelengths where microlensing should be less of an
issue. The fact that the image flux ratios seem to be similar in both the X-ray and the IR
bands would be consistent with this interpretation, although the reasonably large
uncertainties (Table \ref{tab_magtab}) mean that strong conclusions cannot be drawn
from the X-ray data here. It is also worth noting, though, that the long-term X-ray
lightcurve (which formally shows the integrated flux, but in reality is likely dominated by
image A) appears to be relatively stable over a baseline of $\sim$8--9 years (in the
observed frame; see Fig. \ref{fig_lc}); there is a hint of a variability event in the XRT data
just prior to MJD 58800, but the uncertainties on the most elevated data points are
extremely large.

Combining all of the data, we have coverage of the $\sim$0.3--40\,keV band in the
observed frame, corresponding to $\sim$1--140\,keV in the rest-frame of \src. The
broadband data reveal a very hard X-ray spectrum at lower energies, similar to that
reported by \cite{Matsuoka18} based on \swift\ data, and strong curvature of the
continuum emission at higher energies (see Figure \ref{fig_spec_all}). This curvature
peaks at a rest-frame energy of $\sim$30\,keV, implying a strong contribution from
reprocessing by Compton-thick material. Indeed, based on detailed modelling of the
X-ray spectrum, we find that this must be the case in order to reproduce the observed
data with sensible parameters for the primary AGN continuum. However, this
reprocessing cannot be associated with the line-of-sight absorber, which must be
Compton-thin; we find $N_{\rm{H,los}} \sim 3-4 \times 10^{23}$\,\pcmsq, depending
on the precise model used for the broader continuum, but in all cases Compton-thick
solutions for the line-of-sight absorption are strongly ruled out by the data. The
Compton-thin nature of the X-ray absorption is qualitatively consistent with the broad
emission lines observed in the optical data (\citealt{Matsuoka18, Glikman21}). While
we have been able to undertake a much more detailed spectral analysis, the
line-of-sight column density found here is broadly similar to that inferred from the
lower S/N \swift\ data (\citealt{Matsuoka18}). Although this column density is large,
and places \src\ as the most obscured lensed quasar observed to date, we are still
lacking any lensed systems for which the line-of-sight column is Compton-thick,
despite the fact that the latest estimates find that this should be the case for
$\sim$30--50\% of all quasars, either based on direct searches for Compton-thick
sources or population synthesis modelling of the cosmic X-ray background (\eg\
\citealt{Lansbury17, Lanzuisi18, Ananna19}). However, we stress that the majority of
the lensed quasars known have been identified through optical imaging (\eg\
\citealt{Lemon18, Lemon19, Khramtsov19, Stern21}), which is heavily biased against
highly obscured systems.

The reprocessed emission observed must therefore be associated with Compton-thick
material located away from our line-of-sight. We test two possible scenarios, and find
the data can be equally well fit with models in which the reprocessing occurs in a
torus-like structure (as invoked in the classic AGN unification model; \eg\
\citealt{AGNunimod}) with $N_{\rm{H,tor}} > 3.1 \times 10^{24}$\,\pcmsq, or occurs in
the accretion disc (as seen in other, unobscured lensed quasars; \eg\ \citealt{Reis14nat,
Reynolds14, Walton15lqso}). In the former case, the covering factor of the
Compton-thick torus is inferred to be fairly large, $\Omega_{\rm{tor}}/4\pi > 0.61$, even
through we cannot be viewing the central source through it. This would imply a viewing
angle of $i \lesssim 50$\deg, assuming the torus is an equatorial
structure.\footnote{Formally this assumes that the torus is relatively smooth, such that
there are no Compton-thin lines of sight through the torus itself. However, for a variety
of reasons, including the variable levels line-of-sight absorption seen in some systems
(\eg\ \citealt{Risaliti02, Markowitz14, Guainazzi16}), it is now generally expected that
the torus is actually clumpy to some degree. Although we do not discuss these fits in
detail, since they provide an extremely similar solution to that presented in Table
\ref{tab_borus}, we stress that we have also tested a clumpy torus model in addition to
\borus{12}. Specifically we investigated the \xclumpy\ model (\citealt{xclumpy}), which
assumes an increasing density of clouds with increasing viewing angle, and we again
find the best fit is provided by a Compon-thick torus that has very large covering factor
(equatorial column of $N_{\rm{H}} \sim 3 \times 10^{24}$\,\pcmsq, angular width of
$>72$\deg) and is viewed at a fairly low inclination of $i < 45$\deg.} In the latter case,
we find a tighter constraint on the inclination, $i < 36$\deg, directly from the disc
reflection model. In both cases, we would therefore infer that there is still a significant
column of neutral gas that obscures the central source away from equatorial lines of
sight, implying that this component may have a more spherical than equatorial
geometry. This may exist in addition to a thicker, more equatorial structure, or it could
even be the primary structure of neutral gas surrounding the central source (if the
reprocessing occurs in the inner accretion disc). In either case, the overall distribution
of neutral gas around \src\ appears to differ from the simplest picture of a purely
equatorial torus.

\subsection{\textit{L}$_{\rm{\bf{X}}}$ vs \textit{L}$_{\rm{\bf{IR}}}$}

Following \cite{Stern15}, we place \src\ in the context of other quasars/AGN, both in
terms of its observed and intrinsic properties, by investigating where it lies in the X-ray
vs IR luminosity plane (specifically, rest-frame 2--10\,keV vs 6$\mu$m). Here, we quote
fluxes integrated over all four of the quasar images, mimicking the results that would be
seen with X-ray and IR surveys that do not have the imaging resolution to separate
them. The total observed flux in the rest-frame 2--10 keV band is $F_{2-10,\rm{obs}} =
1.8 \pm 0.7 \times 10^{-13}$\,\ergpcmsqps, corresponding to a luminosity of
$L_{2-10,\rm{obs}} = 9 \pm 3 \times 10^{45}$\,\ergps\ for a luminosity distance of
$D_{\rm{L}} = 2 \times 10^{4}$\,Mpc. Correcting the X-ray flux for the line-of-sight
absorption, we find $L_{2-10,\rm{unabs}} = (2.7 \pm 0.7) \times 10^{46}$\,\ergps\
(as the two preferred models considered here both have similar column densities, there
is good agreement between them regarding the absorption correction, and the
uncertainty range given here represents their combined uncertainty range). Finally, we
de-magnify the flux to compute the intrinsic 2--10\,keV luminosity, $L_{\rm{2-10,int}}$.
At the time of writing, \cite{Glikman21} present two estimates for the total magnification:
$\mu_{\rm{mag}} = 53 \pm 5$, based purely on their current lens modelling, and
$\mu_{\rm{mag}} = 122 \pm 26$, based on a combination of their lens modelling and
the image flux ratios observed in the IR by \hst. An independent lens modelling by
Schmidt et al. (\textit{submitted}) that combines these IR data with more recent UV
data from \hst\ also finds a magnification factor of $\mu_{\rm{mag}} \sim 51$, in good
agreement with the equivalent analysis in \cite{Glikman21}. Nevertheless, give the
contrasting values reported by \cite{Glikman21}, to be conservative we present
estimates for $L_{\rm{2-10,int}}$ (and various quantities derived subsequently) for
both magnification factors (see Table \ref{tab_magtab}). We find $L_{2-10,\rm{int}}
\sim$ \intLx\,\ergps. Similarly, the total observed luminosity at $6 \mu$m is
$L_{\rm 6,obs} = 4.6 \times 10^{47}$\,\ergps\ based on interpolating the {\it WISE}
photometry, and the intrinsic luminosity after correcting for the magnification is
$L_{6,\rm{int}} \sim$ \intLsix\,\ergps.

\begin{table}
  \caption{Derived X-ray/IR luminosities and related quantities for \src, given for both of
  the potential magnification factors reported by \citet{Glikman21}.}
\begin{center}
\begin{tabular}{c c c c}
\hline
\hline
\\[-0.25cm]
Quantity & Unit & $\mu_{\rm{mag}} = 53$ & $\mu_{\rm{mag}} = 122$ \\
\\[-0.25cm]
\hline
\hline
\\[-0.15cm]
$L_{\rm{2-10,int}}$ & [$10^{44}$\,\ergps] & $5.1 \pm 1.4$ & $2.2 \pm 0.7$ \\
\\[-0.3cm]
$L_{\rm{6,int}}$ & [$10^{45}$\,\ergps] & $8.7 \pm 0.8$ & $3.8 \pm 0.8$ \\
\\[-0.3cm]
$L_{\rm{15,int}}$ & [$10^{45}$\,\ergps] & $11.0 \pm 1.0$ & $4.8 \pm 1.0$ \\
\\[-0.3cm]
$L_{\rm{bol,int}}$\tmark[a] & [$10^{46}$\,\ergps] & $9.4 \pm 1.4$ & $4.1 \pm 1.0$ \\
\\[-0.3cm]
$M_{\rm{BH}}$\tmark[b] & [$10^{9}$\,\msun] & $8^{+2}_{-3}$ & $5^{+1}_{-2}$ \\
\\[-0.3cm]
\eddrat\ & & $0.09^{+0.04}_{-0.03}$ & $0.06^{+0.03}_{-0.02}$ \\
\\[-0.3cm]
$\Omega_{\rm{tor,pred}}$\tmark[c] & [$4\pi$] & 0.0 ($<$0.25) & 0.15 ($<$0.35) \\
\\[-0.2cm]
\hline
\hline
\end{tabular}
\label{tab_magtab}
\end{center}
$^{a}$ Assuming a bolometric correction of $8.5 \pm 1.0$ for $L_{\rm{15,int}}$ from
\cite{Runnoe12}. \\
$^{b}$ From \cite{Matsuoka18}, based on \Ha\ and \Hb\ line widths; quoted values are the
averages of the individual \Ha\ and \Hb\ estimates, and their uncertainties correspond to
the full ranges given in \cite{Matsuoka18}, incorporating both \Ha\ and \Hb, and their
quoted statistical uncertainties. \\
$^{c}$ Based on \cite{Brightman15} and $L_{\rm{2-10,int}}$; the full uncertainty range is
consistent with zero in both cases, so we quote both the predicted value and its upper
limit (with the latter in parentheses). \\
\end{table}

\begin{figure}
\begin{center}
\hspace*{-0.35cm}
\rotatebox{0}{
{\includegraphics[width=247.5pt]{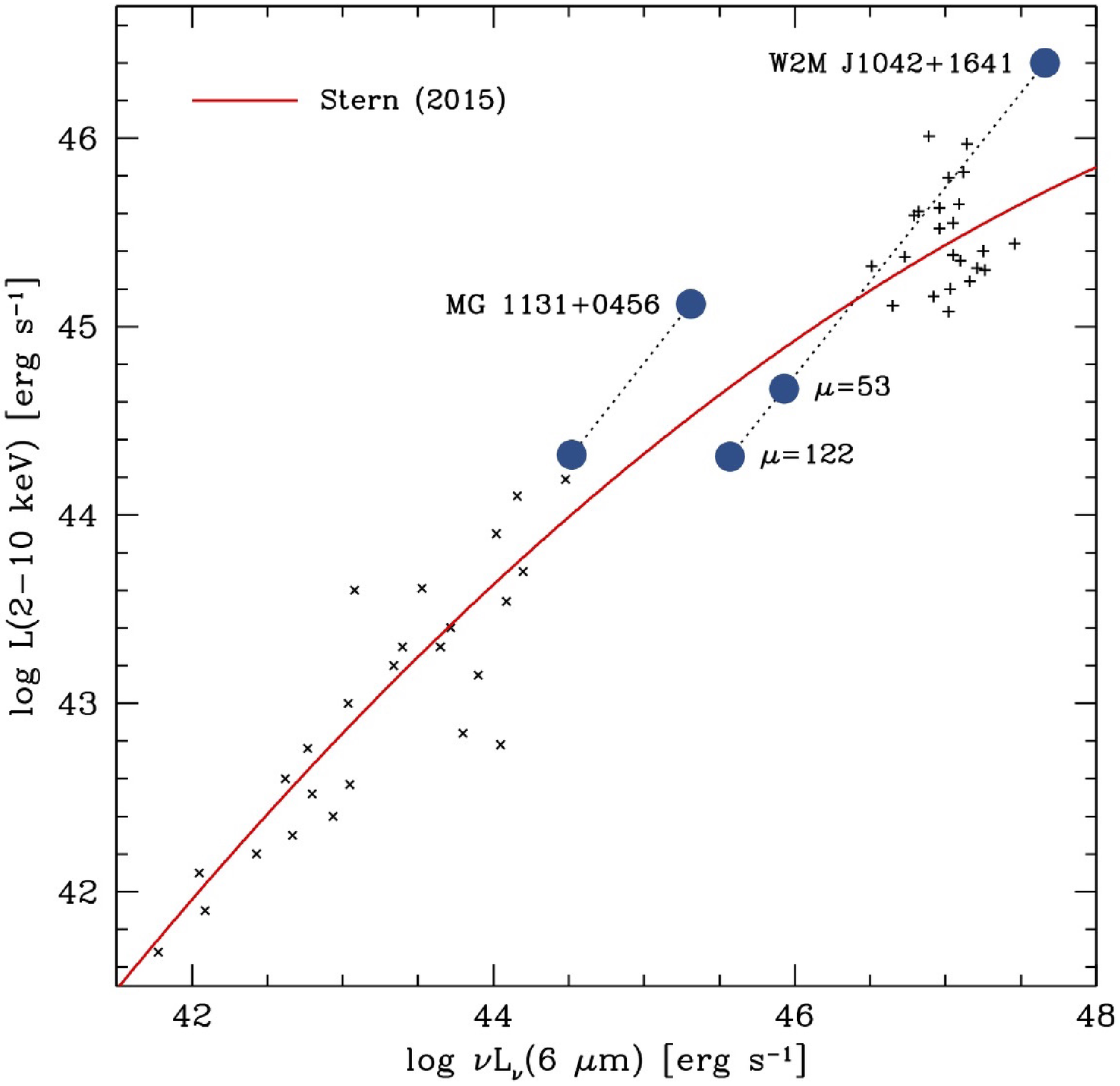}}
}
\end{center}
\vspace*{-0.45cm}
\caption{
Rest-frame, absorption corrected 2--10\,keV X-ray luminosity against rest-frame
6$\mu$m luminosity for the sample of non-lensed AGN compiled by \citet{Stern15}.
These cover a wide range luminosity; local Seyfert galaxies from \citet{Horst08} and
\citet{Gandhi09} are shown with $\times$ symbols, while luminous quasars from
\citet{Just07} are shown with $+$ symbols. The best-fit relation between the X-ray and
IR luminosities derived by \citet{Stern15}, which is linear at lower luminosities and
begins to flatten at higher luminosities, is shown with the solid red curve. Large circles
show the results for \src\ both before and after the lensing magnification is accounted
for. Intrinsically, \src\ lies close to observed relation, at a luminosity where the trend is
roughly linear. However, the lensed values would appear to place this source far above
the relation. This is similar to the lensed radio galaxy MG\,1131+0456 (also shown;
\citealt{Stern20}), and may suggest a new method to identify lensed AGN candidates
from wide-field X-ray and IR surveys.}
\label{fig_XrayIR}
\end{figure}

These values are shown in Figure \ref{fig_XrayIR}, along with a sample of non-lensed
AGN from which \cite{Stern15} compute their best fit relation between the 2--10\,keV
and 6\,$\mu$m luminosities (note that these X-ray luminosities are also all corrected for
line-of-sight absorption). While this is roughly linear at lower luminosities, \cite{Stern15}
find that the trend begins to flatten at higher luminosities, such that an equivalent
increase in IR luminosity would gradually correspond to a smaller and smaller increase
in X-ray luminosity. Analysing the Einstein ring MG\,1131+0456, \cite{Stern20} speculate
that this might provide an opportunity to identify new lensed quasar candidates by
combining wide-field X-ray and infrared surveys (\eg\ \erosita\ and \wise;
\citealt{EROSITA_tmp, WISE}), as sources that intrinsically lie on the linear part of the
relation, when significantly magnified, would appear to have anomalously high X-ray
luminosities (see also \citealt{Connor22}). The results found here for \src\ would also
suggest this could be a promising method; given its IR luminosity, \src\ would indeed
appear to have an unusually high X-ray luminosity, but after accounting for the
magnification, lies much closer to the relation for unlensed AGN. It should be stressed
that this would also require significant efforts to obtain source redshifts, which would be
a critical part of such an identification method. Nevertheless, this approach might be
particularly  appealing, as it should be most sensitive to the rare quadruply lensed
systems, which have the strongest total magnification.

We note that these comparisons have assumed that the lensing magnifications are the
same in the IR and X-ray bands. The fact that this does bring \src\ into better agreement
with the results seen from unlensed quasars suggests that this assumption is not
unreasonable. This assumption is also made in the other pilot studies of lensed quasars
that have explored the idea of pre-selecting such sources based on their X-ray vs IR
properties (\citealt{Stern20, Connor22}). Indeed, if we take the lens model for \src\
presented by \cite{Glikman21}, we find that any differential magnification should be at
less than the 10\% level for emitting regions smaller than 10\,pc. This lens model
corresponds to the magnification factor of $\mu = 53$, and assuming that the
characteristic size scale of the torus is set by the dust sublimation radius,
$R_{\rm{sub}}$, based on the expression outlined in \cite{Barvainis87} we find
$R_{\rm{sub}} \sim 4$\,pc for this magnification factor (assuming a typical grain size of
0.05\,$\mu$m and a sublimation temperature of 1500\,K and conservatively taking
$L_{\rm{UV}} \sim L_{\rm{bol}}$). As such, assuming a common magnification factor for
the X-ray and the IR data is likely a reasonable approach here. It is also worth noting that
the ratio of the luminosity of the narrow iron K$\alpha$ emission -- assumed to be
associated with distant reprocessing -- to the absorption-corrected 10--50\,keV
luminosity is $L_{\rm{FeK}}/L_{10-50} = 3.2^{+1.1}_{-1.8} \times 10^{-3}$, assuming the
two experience the same magnification; although the uncertainties are relatively large,
this is extremely similar to the typical ratio seen for unlensed quasars ($\sim$3--4
$\times$ 10$^{-3}$; \citealt{Ricci14feK}).

In general, though, it may be plausible that the X-ray and IR magnification factors are
not the same if there is significant microlensing from structure in the lens, given that at
least in the case of unobscured quasars the X-ray emission is expected to come from
much smaller scales than the IR emission. Should this occur, the general expectation is
that the more compact X-ray source would experience a larger total magnification
(\eg\ \citealt{Chartas12, Hutsemekers21}). As long as the sources are intrinsically close
to the X-ray vs IR trend reported in \cite{Stern15}, though, then enhanced X-ray
magnification (vs the IR) would actually drive sources even further away from this trend
and make them appear as even more extreme outliers. This method of selecting lensed
quasar candidates would therefore still be suitable even in this scenario.

\subsection{Reprocessing and Black Hole Growth}

In principle, we may also be able to use the intrinsic 2--10\,keV luminosity to estimate
the expected covering factor for the Compton-thick phase of the surrounding medium,
assuming that the anti-correlation between these quantities seen in local AGN
(\citealt{Brightman15}) also holds for \src. For the above luminosity, we find a predicted
covering factor of $\Omega_{\rm{tor, pred}}/4\pi <$ \predFcov\ (Table \ref{tab_magtab}).
If this trend does hold, we would therefore expect only a small contribution from any
Compton-thick torus in terms of reprocessed emission. This would in turn suggest that,
of the two scenarios considered for the dominant source of the reprocessed emission 
(torus vs disc reflection), the disc reflection interpretation may be the more likely
solution. We stress, though, that it is not clear whether the \cite{Brightman15} relation is
still relevant here, as there is good evidence that the fraction of `obscured' AGN (those
with $N_{\rm{H}} > 10^{22}$\,\pcmsq) increases with increasing redshift for a given
X-ray luminosity (\eg\ \citealt{Ueda14, Buchner15}). The relevant issue here, though, is
whether a similar trend is also present for Compton thick AGN specifically. There is
conflicting evidence over this point (\citealt{Brightman12} do find evidence for a
qualitatively similar trend for Compton-thick AGN, while \citealt{Buchner15} find the
fraction of Compton-thick AGN to be constant with redshift), but if a similar trend is
present then the \cite{Brightman15} trend would not formally be suitable for use with
\src. We also note again the contrasting magnification factors reported for this source.

Nevertheless, the nominal difference between the inferred covering factor for the 
Compton-thick phase of the torus and the best-fit prediction for even the lower
intrinsic luminosity estimate (i.e. the one associated with $\mu_{\rm{mag}} = 122$)
based on \cite{Brightman15} is pretty large. If the disc reflection model is the more
appropriate solution, future efforts to obtain tighter constraints on the strength of the
reflection may be of significant interest. Although the level of obscuration will make
direct constraints from broad Fe K emission difficult, the strength of the reflection can
still potentially provide information on the spin of the black hole if a
reflection-dominated scenario can be confirmed (\eg\ \citealt{Dauser14}). If the
potential variability event noted earlier is real, and intrinsic to the source (as opposed
to driven by microlensing), then based on the masses reported in \cite{Matsuoka18}
and the observed-frame duration of $\sim$15 days light-crossing arguments would
suggest the majority of the X-ray flux comes from a region of less than
$\sim$10--15\,\rg\ in size (broadly similar to the X-ray source sizes inferred in other
lensed quasars where microlensing constraints can be placed, \eg\ \citealt{Dai10, 
MacLeod15}). This may imply a mild preference for the disc reflection scenario, but
given the large uncertainties regarding the significance of this event, also highlighted
above, we strongly caution against over-interpretation here. Given the observed flux,
improved constraints on the reflection may be challenging with \nustar, requiring
extremely deep exposures, but \src\ would likely be an excellent target for the next
generation of hard X-ray observatory (\eg\ \hexp; \citealt{HEXP_tmp}).

In order to estimate the intrinsic bolometric luminosity of \src, we also extract the
15$\mu$m luminosity following the same methodology as for the 6$\mu$m luminosity,
and combine this with the 15$\mu$m bolometric correction reported by \cite{Runnoe12}:
$\kappa_{15} = 8.5 \pm 1.0$ (such that $L_{\rm{bol,int}} = \kappa_{15} L_{\rm{15,int}}$;
see also \citealt{Richards06sed}). After accounting for the two possible magnification
factors considered here, we find that $L_{\rm{15,int}}$ = \intLftn\,\ergps, implying
$L_{\rm{bol,int}}$ = \intLbol\,\ergps. These values are in excellent agreement with the
bolometric luminosities reported by \cite{Matsuoka18} based on the 5100\,\AA\
luminosity. \footnote{Here we choose to present the `traditional' form of the
bolometric correction presented by \cite{Runnoe12}, i.e. $L_{\rm{bol}} = 8.5
L_{\rm{15}}$. We do note, however, that \cite{Runnoe12} also discuss a slightly more
complex relation of the form $\log(L_{\rm{bol}}) = 10.514 + 0.787 \log(L_{15})$, which
they argue gives a mildly better fit to the data (at just below 98\% confidence); this
would reduce the $L_{\rm{bol,int}}$ estimates presented here by a factor of $\sim$1.5.
We choose to present the more traditional approach because this gives a slightly better
agreement with the bolometric luminosities based on the 5100\,\AA\ data presented
by \cite{Matsuoka18}, but stress that both of the 15\,$\mu$m bolometric conversions
presented by \cite{Runnoe12} would give good agreement with \cite{Matsuoka18}.}
For the masses also reported in \cite{Matsuoka18}, based on the \Ha\ and \Hb\ line
widths ($M_{\rm{BH}} \sim 5-8 \times 10^{9}$\,\msun), these bolometric luminosities
would  correspond to fairly modest Eddington ratios of $\lambda_{\rm{E}} =
L_{\rm{bol,int}}/L_{\rm{E}} \sim 0.05-0.1$. Given the intrinsic X-ray luminosities
calculated  earlier, these bolometric luminosities would also imply a large 2-10\,keV
bolometric correction of $\kappa_{2-10} \sim 200$ is appropriate for \src. In the local
universe, a bolometric correction of $\kappa_{2-10} \sim 200$ would be unusually high
for such modest Eddington ratios (\citealt{Vasudevan09, Lusso10}). However, such a
correction would be broadly consistent with the connection between $\kappa_{2-10}$
and $L_{\rm{bol}}$ reported by \cite{Marconi04}, which would predict $\kappa_{2-10}
\sim 100$.

Regardless, for the Eddington ratio of \eddrat\ $\sim 0.05-0.1$ inferred here, the current
luminosity of \src\ would probably not be sufficient to drive away the neutral gas and dust
along our line-of-sight and bring about a classic unobscured quasar phase, assuming a
standard gas-to-dust ratio. In terms of the \eddrat--\nh\ plane discussed frequently in the
literature (\eg\ \citealt{Fabian08, Ishibashi15, Ricci17nat}), \src\ currently resides in the
long-lived absorption regime, computed by considering the expected radiation pressure
on the accompanying dust. This is in contrast to other populations of red, obscured
quasars (\eg\ \citealt{Banerji14, Lansbury20}), and may have interesting implications if
\cite{Matsuoka18} are correct about the high $M_{\rm{BH}}/M_{\rm{gal}}$ ratio compared
to other similar systems. Presumably this would imply that the black hole has already
undergone a significant (and perhaps abnormal) amount of growth, and yet it remains
buried behind a significant column of gas and dust. Either this growth was somehow
unable to blow out the obscuring medium, in contrast to general expectations for quasar
evolution (\eg\ \citealt{Hopkins08}), or this has somehow been replenished after the last
major quasar phase in \src.

\section*{ACKNOWLEDGEMENTS}

The authors would like to thank the reviewer for their detailed feedback, which helped
to improve the final version of the manuscript.
The authors would also like to thank T. Schmidt and T. Treu for providing advance details
of their lens modelling.
DJW acknowledges support from the Science and Technology Facilities Council
(STFC) in the form of an Ernest Rutherford Fellowship (grant ST/N004027/1).
The scientific results reported in this article are based in part on observations made by
the NASA's \chandra\ X-ray Observatory, as well as data obtained with \xmm, an ESA
science mission with instruments and contributions directly funded by ESA Member
States.
This research has also made use of data obtained with \nustar, a project led by Caltech,
funded by NASA and managed by the NASA Jet Propulsion Laboratory (JPL), and has
utilized the \nustardas\ software package, jointly developed by the Space Science
Data Centre (SSDC; Italy) and Caltech (USA).
For the purpose of open access, the author(s) has applied a Creative Commons
Attribution (CC BY) licence to any Author Accepted Manuscript version arising.


\section*{Data Availability}

All of the raw observational data utilized in this article are publicly available from ESA's
\xmm\ Science Archive,\footnote{https://www.cosmos.esa.int/web/xmm-newton/xsa}
NASA's HEASARC archive,\footnote{https://heasarc.gsfc.nasa.gov/} and NASA's
\chandra\ Data Archive.\footnote{https://cxc.harvard.edu/cda/}

\bibliographystyle{/Users/dwalton/papers/mnras}

\bibliography{/Users/dwalton/papers/references}

\label{lastpage}

\end{document}